\documentclass[epj]{svjour}
\usepackage{graphicx,amsmath,amssymb}

\pdfoutput=1

\newcommand{\be}{\begin{equation}}
\newcommand{\ee}{\end{equation}}
\newcommand{\bea}{\vspace{0.25cm}\begin{eqnarray}}
\newcommand{\eea}{\end{eqnarray}}

\def\PLA{{Phys. Lett.}  A }

\def\PRA{{Phys. Rev.} A }

\begin{document}
\title{Experimental local realism tests without fair sampling assumption}
\author{G. Brida, M. Genovese, F.Piacentini}
\institute{ I.N.RI.M., Strada delle Cacce 91, 10135 Torino, Italia
\and \email{genovese@inrim.it} }


\abstract{ Following the theoretical suggestion of Ref.
\cite{sant1,sant2}, we present experimental results addressed to
test restricted families of local realistic models, but without
relying on the fair sampling assumption. }

\maketitle 
\section{Introduction}

 The quest for a conclusive test of local realism has now a
pluridecennial history \cite{prep}, beginning with the celebrated
1965 Bell's paper \cite{bell} demonstrating that every Local Hidden
Variable Theory (LHVT) cannot reproduce all the results of quantum
mechanics when correlation among measurements on entangled states
are considered. In particular, some inequalities always satisfied by
every LHVT can be violated by Standard Quantum Mechanics (SQM).

Since then many experiments have been realized for testing Bell
inequalities \cite{prep},  with a recent intensification due to the
application to the emerging field of quantum information.
Nevertheless, whilst a clear spacelike separation has been
unequivocally achieved, still a conclusive test is missing, because
of low detection efficiency that requires the additional hypothesis
that the observed sample is a faithful representation of the whole
one (the so-called \textit{fair sampling} assumption, leading to the
detection loophole).

In the last years relevant progresses for eliminating this loophole
have been realized \cite{prep}, both based on improvements of
detectors and on the result that for non-maximally entangled states
the limit for an elimination of this problem is lowered from $82 \%$
to $66.7 \%$ \cite{eb,nosb}. However, a conclusive experiment is
still missing.

On the other hand, specific local realistic models built for
surviving traditional Bell inequalities tests, as stochastic optics
\cite{stocop} (a branch of stochastic elecrodynamics \cite{sed}),
have been falsified by specifically addressed experimental tests
\cite{prep,spuc,piomb}.

Recently, some works appeared \cite{sant1,sant2} with the purpose to
propose some specific inequality verified by a restricted families
of LHVT, but whose test, realizable by exploiting polarization
entangled photon pairs generated by parametric down-conversion
(PDC), would be free from the detection loophole. The aim of this
paper is to present an experimental test of these theoretical
proposals.

\section{Theoretical motivations}

\par
In  \cite{sant1,sant2}  a set-up based on polarization bipartite
entangled states of photons is considered, where one measures the
single count rates $R_1$ and $R_2$ on the idler and signal arm, the
coincidence rate $R_{12}$ and the total pair production rate $R_0$.
One has
\begin{equation}\label{rates}
    R_1=R_2=\frac{1}{2}\eta R_0 ~~~~~~~~~~ R_{12}=\frac{1}{4}\eta^2R_0[1+V\cos(2\phi)]
\end{equation}
where $\phi$ is the angle between the two polarization axes, $V$ the
observed visibility and $\eta$ the  detection
efficiency  (considered identical for the two arms).\\

The results of such an experiment can be reproduced by a LHV model
constituted by the functions \cite{sant1}:
\begin{equation}\label{LHV1}
   \rho(|\lambda_1-\lambda_2|)\geq0~~~~~~0\leq
   P_j(|\lambda_j-\phi_j|)\leq1~~~~~~~~~j=1,2$$
   $$\int\rho(|\lambda_1-\lambda_2|)d\lambda_1d\lambda_2=1
\end{equation}
where $\lambda_1,\lambda_2$ are the hidden (angular) variables and
$\phi_1,\phi_2$ the angles of the polarization planes with respect
to an axis perpendicular to pump direction and polarization. The
predictions for the single count and coincidence rates are:
\begin{eqnarray}
  \frac{R_j}{R_0}=\int\rho(|\lambda_1-\lambda_2|)P_j(|\lambda_j-\phi_j|)d\lambda_1d\lambda_2 \\
  \frac{R_{12}}{R_0}=\int\rho(|\lambda_1-\lambda_2|)\prod_{j=1}^2P_j(|\lambda_j-\phi_j|)d\lambda_1d\lambda_2
\end{eqnarray}
\par
From this model, introducing the visibility parameters:
\begin{equation}\label{VAVB}
   V_A=\frac{R_{12}(0)-R_{12}(\pi/2)}{R_{12}(0)+R_{12}(\pi/2)}$$$$
   V_B=\sqrt{2}\frac{R_{12}(\pi/8)-R_{12}(3\pi/8)}{R_{12}(\pi/8)+R_{12}(3\pi/8)}
\end{equation}
one derives the inequality \cite{sant1}:
\begin{equation}\label{santos1}
   \frac{V_B}{V_A} \geq F=1+\cos^2\left(\frac{\pi\eta}{2}\right)\left[V_B-\frac{\sin^2(\pi\eta/2)}{(\pi\eta/2)^2}\right]
\end{equation}
that can be violated in quantum mechanics (where, in the ideal case,
$V_B/V_A=1$). An accurate test of this inequality will be the first
achievement of our experiment \cite{ops}.
\par
Then we consider another  inequality of the same type, but valid for
a larger family of LHVT \cite{sant2}. In this case, one takes into
account a set of four hidden variables
$\lambda\equiv\{\chi_1,\chi_2,\mu_1,\mu_2\}$ (of two different
types) instead of two. The equations of this local realistic model
are \cite{sant2}:
\begin{equation}\label{LHV2}
   \rho(\lambda)=\rho_{\chi}(|\chi_1-\chi_2|)g_1(\mu_1)g_2(\mu_2)\geq0$$$$
   \int\rho_{\chi}(|\chi_1-\chi_2|)d\chi_1d\chi_2=\int
   g_j(\mu_j)d\mu_j=1
   $$$$
   0\leq P_j(|\chi_j-\phi_j|)=\int Q_j(\mu_j,|\chi_j-\phi_j|)g_j(\mu_j)d\mu_j\leq1
\end{equation}
and give the expected ratios (after integrating over the $\mu_j$
variables):
\begin{eqnarray}
  \frac{R_j}{R_0}=\int\rho_{\chi}(|\chi_1-\chi_2|)P_j(|\chi_j-\phi_j|)d\chi_1d\chi_2 \\
  \frac{R_{12}}{R_0}=\int\rho_{\chi}(|\chi_1-\chi_2|)\prod_{j=1}^2P_j(|\chi_j-\phi_j|)d\chi_1d\chi_2
\end{eqnarray}
\par
Let us consider a system with rotational invariance, in which the
coincidence rates depend only on the relative polarization
orientation $\phi=\phi_1-\phi_2$, and put the probabilities
$P_1=P_2=P$ (perfect symmetry between the apparatuses on the two
branches of PDC). By introducing the parameter (where $()_+$ means
putting zero if the quantity inside the bracket is negative):
\begin{equation}\label{D_eta}
    D(\eta)\equiv\frac{4}{3\pi}\sqrt{\frac{2}{3\eta}-\frac{1}{2}-\frac{\sin^4(\pi\eta/2)}{(\pi\eta/2)^4}}
    \left(V-\frac{\sin^2(\pi\eta/2)}{(\pi\eta/2)^2}\right)^\frac{3}{2}_+
\end{equation}
with:
\begin{equation}\label{V}
    V=2\frac{\sum_{j=1}^n R_{12}(\phi_j)\cos(2\phi_j)}{\sum_{j=1}^n R_{12}(\phi_j)}
\end{equation}
one derives the following inequality:
\begin{equation}\label{santos2}
    \Delta_{min}=\left[\frac{n\sum_{j=1}^n R^2_{12}(\phi_j)}{\left(\sum_{j=1}^n R_{12}(\phi_j)\right)^2}\right.-$$$$-2\left.\frac{\left(\sum_{j=1}^n R_{12}(\phi_j)\cos(2\phi_j)\right)^2}
    {\left(\sum_{j=1}^n R_{12}(\phi_j)\right)^2}-1\right]^{(1/2)}\geq
    D(\eta)
\end{equation}
where  $\phi_j=\pi\cdot (j-1)/n$.

Testing this second inequality will be the second purpose of the
present work.

\section{Experimental set-up and results}
\par
\begin{figure}[htbp]
\includegraphics[width=300px]{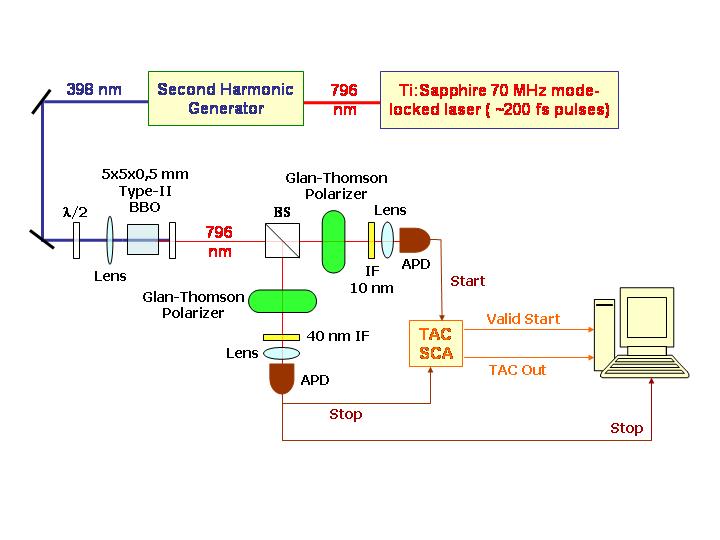}
\caption{Scheme of the experimental setup.} \label{setup}
\end{figure}

In order to test these inequalities, we measured the variation of
the coincidence rate $R_{12}(\phi)$ with respect to the variable
$\phi$, using the experimental setup reported in Fig. \ref{setup}.

In our set-up a 0.2 W, 398 nm pulsed (with 200 fs pulses) laser
beam, obtained by second harmonic generation from a
titanium-sapphire beam at 796 nm, pumps a type-II BBO 5x5x0,5 mm
crystal. With a proper selection of the phase-matching conditions,
we obtain the collinear degenerate (796 nm) PDC and send it to a
non-polarizing 50\%-50\% beam splitter (BS). Passing through the BS,
the state changes like this:
\begin{equation}\label{HV-BS}
    |HV\rangle\rightarrow\frac{1}{2}\Big(|H_rV_r\rangle+|H_tV_t\rangle+|V_rH_t\rangle+|H_rV_t\rangle\Big)
\end{equation}
where the pedices $r$ and $t$ respectively indicate the reflection
and transmission directions.\\
By postselecting the cases when photons follow different paths, one
obtains the Bell state:
\begin{equation}\label{BS-Bell}
    |\psi_+\rangle=\frac{1}{\sqrt{2}}\Big(|V_rH_t\rangle+|H_rV_t\rangle\Big)
\end{equation}
On the transmission ($t$) arm we put a Glan-Thomson polarizer, a
pinhole, an interference filter (FWHM 10 nm, $\lambda_{peak}$ 795
nm) and an optical fiber coupler connected with a multi-mode fiber
to the detector, a Perkin-Elmer SPCM-AQR-14-FC-9491 silicon
avalanche photo-diode (quantum efficiency at 796 nm: 62\%); the
reflection ($r$) arm featured a Glan-Thomson polarizer, a large band
interference filter (FWHM 40 nm, $\lambda_{peak}$ 800 nm) and the
same fiber coupling and detection devices as in the other arm.
Taking the $t$ photons as trigger, a coincidence window is opened on
the arm $r$ detector to reveal the correlated pairs: this is
obtained by sending the output of the $t$ detector as start to a
Time-to-Amplitude Converter (TAC) that receives the $r$ detector
signal as stop. The 20 ns coincidence window is set properly to
exclude spurious coincidences with PDC photons of the following
pulse (we remind that the repetition rate of the laser is 70 MHz).
The TAC output is then acquired via PC, together with the TAC's
Valid Start counts (giving us the total number of opened coincidence
windows). The background has been evaluated and subtracted by
measuring the TAC output of the system without PDC generation
(simply rotating the pump polarization by 90° before the nonlinear
crystal).
\par
\begin{figure}[htbp]
\includegraphics[width=300px]{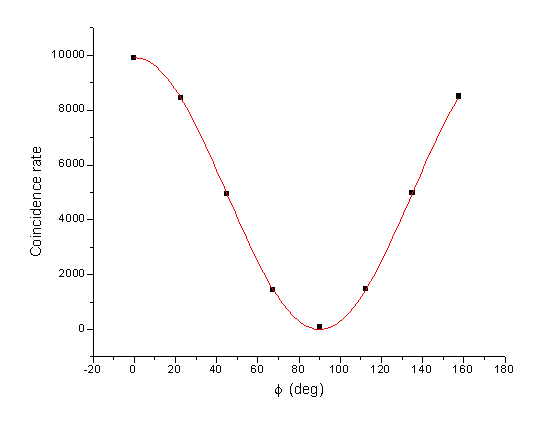}
\caption{Coincidence rates versus $\phi$ (angle between the axes of
$t$ and $r$ polarizers). The experimental data are compared with the
theoretical curve $\cos^2(\phi)/2$, with whom are in perfect
agreement (correlation coefficient $r=0.9998$). Uncertainty bars are
too small for being visible.} \label{BELLcurve}
\end{figure}

To violate the inequalities presented in the previous section, we
fixed the polarization axis of both the Glan-Thomson polarizers at
an angle of $\frac{\pi}{4}$ with the horizontal plane, and then we
performed a scan of the coincidence rates $R(\phi)$ (where $\phi$
represents the difference between the polarization planes) simply
rotating the polarizer on the $r$ arm by $\frac{\pi}{8}$ per step,
until we cover the whole interval $\phi\in[0,\pi[$: the rate
obtained for each step is the average of 30 different
acquisitions of 30 s each.\\
The obtained curve is shown in fig. \ref{BELLcurve}.\\

As a first test for our entangled state, we checked (of course
within the fair sampling hypothesis) the Clauser-Horne inequality
\cite{prep} , written in the form:
\begin{equation}\label{CH}
\frac{\big|R(\frac{\pi}{8})-R(\frac{3\pi}{8})\big|}{R_{tot}}\leq\frac{1}{4}
\end{equation}
The collected data, with a measured visibility $97.8 \%$, led to
the following violation:
\\\\
\begin{tabular}{|c|c|c|}
  \hline
  CH violation & $\sigma$ & $\frac{violation}{\sigma}$
  \\
  \hline
  0,1026 & 0,0021 & 47,8\\
  \hline
  \label{table1}
\end{tabular}
\\\\
Then, to violate the inequality \ref{santos1}, we evaluated the
quantities:
\\\\
\begin{tabular}{|c|c|}
  \hline
  $V_B$ & $V_A$
  \\
  \hline
  0,9985 $\pm$ 0,0030 & 0,9784 $\pm$ 0,0017
  \\
  \hline
  \hline
  $V_B/V_A$ & $F$
  \\
  \hline
  1,0205 $\pm$ 0,0048 & 1,0876 $\pm$ 0,0009
  \\
  \hline
  \label{table2}
\end{tabular}
where, following the indications of Ref. \cite{sant2}, we used for
the value of the quantum efficiency the nominal detector efficiency
as reported on the manufacturer data sheet for our wave length,
$\eta= 62 \%$.

Of course, the difference of $V_A,V_B$ from unit and from each other
derives from unavoidable imperfections of the experimental
apparatus; nevertheless a clear violation of the inequality
\ref{santos1} is observed:
\\\\
\begin{tabular}{|c|c|c|}
  \hline
  $F-V_B/V_A$ & $\sigma$ & $\frac{violation}{\sigma}$
  \\
  \hline
  0,0671 & 0,0057 & 11,7
  \\
  \hline
  \label{table3}
\end{tabular}
\\\\
Finally, we present the results for the inequality (\ref{santos2})
(with again $\eta= 62 \%$). Our data led to a violation:
\\\\
\begin{tabular}{|c|c|c|}
  \hline
  $D(\eta)-\Delta_{min}$ & $\sigma$ & $\frac{violation}{\sigma}$
  \\
  \hline
  0,0073 & 0,0022 & 3,3
  \\
  \hline
  \label{table4}
\end{tabular}
\\\\For the sake of completeness,
we mention that both the inequalities (\ref{santos1},\ref{santos2})
still are violated even without background subtraction (of 11.6 and
1.7 standard deviations respectively).

Thus, we have clearly shown a violation of the predictions made upon
the LHVT models presented in \cite{sant1} and \cite{sant2},
whilst perfectly agreeing with SQM.\\

\par
\section{Conclusions}
In summary, we have experimentally demonstrated, following the
theoretical predictions of \cite{sant1,sant2}, that the specific
families of hidden variable theories considered there fail to
properly reproduce the observed correlations between measurements of
bipartite polarization photon entangled states, and are therefore falsified by our experiment.\\\\

\begin{acknowledgement}
This work has been supported by MIUR (PRIN 2005023443-002), by
Regione Piemonte (E14) and by grant RFBR-Piedmont 07-02-91581-ASP.

We acknowledge fruitful discussions with E. Santos.
\end{acknowledgement}

\newpage

\end{document}